# Software Estimation's Risk in Pakistan Software Industry

Suresh Kumar[1], Qaisar Imtiaz[2], Sarmad Mahar[3]

Department of Software Engineering,

PAF - Karachi Institute of Economics & Technology Pakistan

Skamrani2002@gmail.com, qaisarimtiaz92@gmail.com, Sarmad.mahar@gmail.com

**Abstract:** Software and IT industry in Pakistan has seen a dramatic growth and success in past few years and is expected to get doubled by 2020, according to a research. Software development life cycle comprises of multiple phases, activities and techniques that can lead to successful projects, and software evaluation is one of the vital and important parts of that. Software estimation can alone be the reason of product's success factor or the product's failure factor. To estimate the right cost, effort and resources is an art. But it is also very important to include the risks that may arise in the in a software project which can affect your estimates. In this paper, we highlight how the risks in Pakistan Software Industry can affect the estimates and how to mitigate them.

## I. Introduction:

Today, the software industry in the Pakistan has emerged as very big and successful industry in Pakistan and is expected to be doubled by 2020, according to a research. There are many big IT and Software companies, which have international clients and very large or successful projects. As the world of Technology and automation is getting bigger, our lives are being surrounded by software and to develop software, we have to go through many phases, activities and techniques.

Software Risk Estimation is one of the most important risks facing the Pakistan Software Industry. The ability to accurately measure software size, effort, and schedule is important and should be considered by the Software industry. There are a number of technical factors, financial and otherwise, which are intrinsically uncertain.

One of most challenging and the success factor of software development is the Software Estimation.

Cost estimates include procedures or methods that help us estimate the actual cost that will be required for our software and are considered one of the most complex and challenging tasks for software companies.

Estimating an entire project is more complex for those persons who are completely not aware with the active weight or acquire a very little software information, so estimator is should be a experienced person of the relevant field that they will able to estimate the project cost and schedule more tends to accuracy.

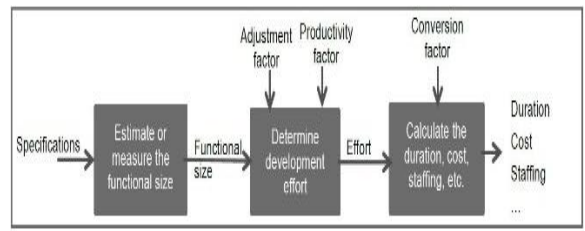

Figure 1 – A common estimation flow for software.

Software estimation provides us the probabilistic idea of;

- Management of project cost
- Completion time for project
- Resources required for project

Performing a successful software project estimation can lead to project success, whereas,

Software estimation can also be the only reason why our projects gets failed or cancelled. While we estimate our project, it is very important to include risks that can fall in during the development cycle. Risk is anything that is unexpected happening in our project. The software's risk exists because the future is uncertain and there are many known and unknown factors that will not be included in the project plans.

While there are some risks that are universal and applied to almost every software project, but in Pakistan Software House industry, there are some extra risks or unwanted events that can happen and affect our estimates.

Software risks means some extra time, effort and cost in project and if that is not included in our estimation plan, it could be the only reason behind inaccurate estimation plan. As in the Pakistani software sector we are still moving away from the planning process, the Gantt charts are emerging and on a more forward-looking path.

Software development is a team work and multi phased job, where risk can be produced due to many reasons at multiple stages. Such risks can affect the software development process or the performance of the people who develop the software. A risk can be a small task, which can cause the delay, re work in some component, hurdle in the process or etc., which can be a cause of software failures. The most unsafe risk is that, Software development projects often run out of time and budget problems and do not meet all the needs of users (Boehm B. , 1981).

Software cost and estimation is considered as a most important and crucial part of the software development. And mostly people neglect the importance of the estimation and the risk that can occur at the estimation process. In this paper, we will categorize and list down the risks, that can arise in the software estimation process and will suggest some mitigation techniques to deal with the risks.

**II. Related Work**

There has been much study about software estimation and how to make it an effective process. But it is also very important to include risks and their probability of occurrence while estimating the project, because if the risks are not measured that can affect the estimates. This is especially important at the beginning of projects when the risk is at a very high level. However, the risk is rare. Accidents detected may ignite a fire or simply evaporate. Risks that were not considered or that could arise suddenly and had a profound effect. (K. Jantzen)

According to PASHS Pakistan the growth of IT & Software exports has been the highest in South Asia over the past three years. Exports of Pakistan and ITeS of Pakistan increased by 71% from June 2013 to June 2016, while exports to India increased by 40.6% and in Sri Lanka by 19.9% at the same time. During FY 2016-17, Pakistan & ITe ITeS export exports have grown faster than India and Sri Lanka. Pakistan's growth rate between FY 2016-17 was 16%, India's growth rate is 8% and Sri Lanka's growth rate is 5%. In all, Pakistan's total IT industry revenue is estimated at $ 3.3 billion a year (Annual Report of Pakistan Software Export Board (PSEB))

Various studies also conclude that the most software project failures are due to budget and schedule overrun. Also, one of the main of project failures is unmanaged risk. (June Verner) We define a risk as a land or property development project or environment, which, if ignored, will increase the likelihood of a project failure. (Klein)

According to a study by The Standish Group International, 44% of software projects are more expensive and last longer than expected. Guessing the effort is more accurate; how the business gets better and where the software project respects the commitment to budget, time and quality. (The Standish Group Report, 2014)

## III. Analysis and Approaches

**Risks related to Software Estimation Process.**

Risks are inevitable part of software development process and can occur at any phase of development. We are focusing on the risks that can occur in the software estimation process. We are categorizing these risks into six categories;

**A. Internal Risks**: The risk that arise from the organization itself are known as Internal Risks. Internal risks are often predictable, so they can be avoided or minimized. Internal risk is usually generated by one (or a combination of) human, technical or physical factors. (Hooman Hoodat)

**B. External Risks**: The risks that are not in control of organization are known as External risks. Many external risks are caused by legal, environmental and political changes. The impact of a major environmental disaster on the organization's supply is an external hazard. (Hooman Hoodat). Some other external risks could be;

- Loss of funds.
- Market crises
- Changing customer mindset about product and priority
- Government acts and new laws (Types of Risks in Software Projects, n.d.)

**C. Schedule Risk**: The factors in a project due to which a project's schedule can slip, are known as schedule risks. Schedule risks not only cause a project failure but can also affect the company's profile and economy. The examples of schedule risks could be; (Types of Risks in Software Projects, n.d.)

- Wrong time estimations
- Failure to identify complex functionalities
- Resources are not managed properly
- Project scope's expansions beyond expectation

**D. Budget Risk**: The risks related to the cost of project are known as Budget risks. (Types of Risks in Software Projects, n.d.). This type of risk can cause the project cost overrun. These risks are occurred because;

- The requirements are not properly understood.
- Project scope expanded or changed.

**E. Operational Risk**: Operational risks are the risks that happen to occur in the organization or the team performing the software development. These risks could be;

- Failure to resolve forward-looking conflicts
- Failure to resolve obligations
- Lack of human resources
- No proper domain training
- Lack of resource planning
- Lack of team communication. (Types of Risks in Software Projects, n.d.)

**F. Technical Risk**: The risks that are related with the technicality, performance, or the functionality of the software are known as Technical Risks. The causes of technical risks are;

- Insufficient knowledge of technology of project.
- Difficult project module integration.
- Continuous requirement changes. (Types of Risks in Software Projects, n.d.)

| No. | Risk | Detail Of Risk | Stakeholder Concerned | Software risk component related to |
|---|---|---|---|---|
| 1 | Staff Shortage | Lack of qualified staff and their replacement | Clients, users, subordinates, suppliers, managers, project manager. | Risk of human resource management |
| 2 | Tough schedules and budgets | Development time and budget are poorly estimated | Customers, bosses, project owner | Scheduling and timing |
| 3 | Creating wrong development architecture | Software development without aware of domain knowledge | Customer, project manager | System functionality |
| 4 | Useless user interface | Unappealing user interface | User, project manager | |
| 5 | Gold plating | Adding unnecessary features to software due to professional interest or pride or user preferences | Sub-ordinates, users, project manager | Requirements management |
| 6 | The broadcast of the necessary changes continues | Uncontrolled and unexpected changes of system functions and features | Sub-ordinates, users, project manager | |
| 7 | Poor architecture of ready packages | Lack of quality in delivered components | Customers, bosses, project manager | Sub-contracting |
| 8 | Shortfalls in externally performed tasks | Poor quality or unpredictable accomplishment of tasks that are performed outside the organization | Customer, bosses, project manager | |
| 9 | Lack of real-time performance | Lack of performance | Users, customer, maintainers, project manager | Resource usage and performance |
| 10 | Lack of computer science capabilities | Failure to use the system due to lack of technical solutions and computer capabilities | Sub-ordinates, users, customers, project manager | |

Table 1 – A major overview of Boehm's Top 10 Risk List and Stakeholder Perspectives Concerned
(Boehm B. W., 1991)

**IV. Risks in Pakistan Software Industry that can affect the estimates**

Like every country, Pakistan is also emerging and climbing heights like every other country in the software and technology field. Our industry is generally adapting the techniques, methods and approaches that international and other markets follow to develop software. For example, we have adapted agile and we have shifted to technologies like AngularJS, Python and slowly shifting to work on technologies like Machine Learning and Neural networks.

But there are risks which are universal and can occur anywhere and there are some risks that can happen in Pakistan Software Industry that can cause a shift or change in our estimates, which can lead to project failure. In this section, we are listing down the risks that can affect our estimates;

**A. People are switching jobs frequently:** We have seen that most of people switch their job very frequently. Generally people link themselves with the organizations not the projects. And due to that, the estimates are affected. Because when the people change, the effort and time required developing the software changes. Suppose, if an experienced personnel has provided the estimates and now he is leaving the job, which would cause the shift in estimates.

**B. Unexpected/Unwanted events or holidays:** We as a nation have been facing the unexpected or unwanted holidays due to extreme political or some other influences. These holidays are not included in our plans because we never have an idea about them. This also can cause the delay in the plans and software delivery, which results the slip in software schedule and budget.

**C. Team members do not have clear idea about estimation techniques:** The idea of estimating the effort is still new to some people. When the team is asked to provide estimates, the resources provide their estimate in one go and in single value. Estimates that are not properly analyzed and given in single value are very risky, because when they are not properly analyzed there is a high chance that those estimates are not realistic and successful.

**D. Lack of guidance or missing role of software estimator:** At every step or in every phase, if there is a guider, you can come out of any problem. In estimation area also. In Pakistan Industry also, there is a no role separately for Software Estimator who can help with the estimation process. Like there is a separate role for Scrum Master, Project Manager or the Team lead, who can help the team to face the unpleasant situation, if there would a separate role of Software Estimator, the estimation process and the estimates would be very realistic and pleasant.

**F. Extra demands from clients:** We provide estimates what is included in the plan. But if there are some extra demand or frequent change requests from client, they are not estimated and that can cause the delay in the delivery.

**G. Lack of user involvement:** Some researches prove that the end user involvement plays one of important roles in system success. (Muneera Bano). If the end user is not available during the project life cycle, that can cause multiple problems during the development for example, they cannot provide feedback on how the project is shaping up (technically known as User Acceptance Testing), if there is any confusion related to the project requirements, they cannot provide feedback on it and much more. These kinds of problem can also affect the estimates as they process gets slow down on such consequences.

**H. Ambiguous requirements:** When a single requirement specification can be interrupted in many ways, that means that the requirements are not clear. And if our need are not clear before the development starts, that can affect our estimate and can be a reason of project failure, all alone. The estimates you provide to achieve a single requirement or user story, only gets successful when the requirements are clear to everyone in the project team. Because if the requirement is ambiguous, the team also could not agree on the provided estimate.

**I. The less confidence on use of estimation process:** Theoretically, there are many studies that prove that how software estimation plays a successful part in the process of software development, but practically, its still somehow executed with the less confidence in some organizations. It is very important to have a common understanding between all team members, that how estimates can impact on the success of project and how productive our software development process could be.

**J. Lack of knowledge and use of estimation tool:** Tools related to Cost estimation for software are available in the market which can help in providing the accurate estimates and reducing the manual work involved in the estimates. But in Pakistan's industry, the exposure to estimation tools is still new and very less. With the help of tool, the estimation process could be a very smooth process and then it can be reinforced, as it would reduce the man's time.

**J. Cultural Difference:**
Mostly organizations in Pakistan has international clients, and working with such clients is whole new dimension as the time zone is different, work environment is change and communication methods are also change. These all factors combined, can also affect the estimates. Working around different cultures and time zones can create multiple complexities and dependencies can cause the delay which can result a change into schedules.

**K. Availability of data:** Technology rapid changes but late adoption of technology in Pakistan due that factor of the availability of the data. In a world full of data, cheap and exciting computers, and advanced algorithms, analytics have become increasingly important for business and government success. In a market where analysts describe competitive profit, data is new gold and data scientists are the new boys and girls of gold. When we developing the local government's projects, we are always missing actual data. Another major difficulty in the software costing process is the availability of the information needed to ensure the accuracy of any proposed models, metrics and applicable size. Many software costing model bases on the data. Data like

a. Project Data
Project data is the existing project's data, which can easily available after some iteration.

b. Historical Data
Historical data is pervious used projects data which could maintain be software houses.

c. Industrial Data
But the real problem is that get the industrial data. Due to late adoption of technology we have very small number of industrial data. Testing and costing on actual industrial data is more accurate than by using others developed countries industrial data.

**L. Unstable Environment:**
In Pakistan, estimation process the usability of the environment also impact on the costing in simple way "What is correct today may not tomorrow!", this shows the frequently modifications and improvements in the development process where most of the model need to modified and sync from time to time. (Abualkishik)

**V. Risk in effort estimation**
According to a study by The Standish Group International, 44% of software projects are more expensive and last longer than expected. This may be the result of incorrect and unreasonable estimates.
 Since risk is a major factor that prevents a software project from being timely and budget-friendly, it is important to pay attention during the evaluation process.
Several ways of dealing with this issue were proposed such as the simple method, Function Point Analysis, COCOMO II, etc. In several cases, Leads, project managers use common and easiest approaches that depend on applying a risk factor to the estimated effort, time, size and cost in order to address with the risk in estimation. The managers address these issues on the bases of their experience. From 1 to 10. That means that if the factor is 1, no risk will occur, and when the cause is x, all risks will occur. This approach is not difficult to manage, but it does not allow determining the outcome of a particular risk because the risk factor is affected by all project risks without specifying a specific risk and evaluating one.

Function point (FPA), provides a set of rules for the efficient operation of a software function. This work product is the result of new software for developing and expanding projects for the next release. It is software, which is transferred to the production application at the start of the project.

Function Point Analysis (FPA) Analysis is a Functional size Measurement method. It evaluates the performance delivered to its users, based on an external user view of performance requirements. It measures a rational view of the app as compared to measuring a physical-made view or an in-house technical view. (Uzzafer, 2013)

COCOMO II repairs the pools quoted in the two previously introduced methods; describes a risk factor that, based on the six risk types, indicates each module to be developed (S. L. Pfleeger, 2005). Each risk should be assigned to one type of risk while the risk can be multiple and therefore assigned to different types, which will make it difficult to adjust their treatment. Alternatives are designed to address project risk in cost estimates such as the emergency measurement model proposed by (Uzzafer, 2013). In this model, contingency is defined as the ratio between the expected project costs and the expected costs due to the high impact of the risk events. Therefore, the contingency model provides a final estimate of human months to minimize the significant impact of risk. This model is standard and independent of cost estimation and risk assessment models, and puts the software at risk, but does not offer a different amount of cost estimates and ignores updated software risk estimates [20].

All cost models contain some form of risk assessment, because cost estimates are a matter of course and objective, they can feed directly on the risk management system. Government agencies call for estimates to be presented with 80 percent certainty.

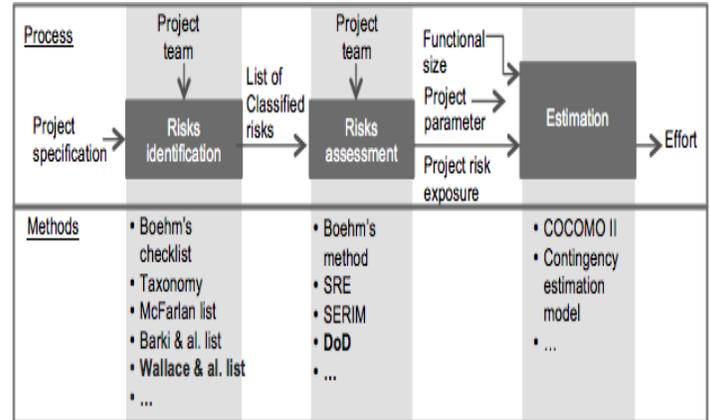

Figure 2 An effort evaluation process that combines risk and examples of support mechanisms.

Our approach aims at this step to assess software risk across the project and to get a complete risk profile of the project.

## VI. Recommendation on Mitigation of Risks

All of the risk analysis activities presented has the sole purpose of assisting the team in developing a risk management strategy. An effective strategy must consider three issues (Pressman):

- Risk avoidance
- Risk monitoring
- Risk management and contingency planning

If the software team uses an effective risk management approach, prevention is always the best strategy. This is achieved by building risk mitigation. We also recommend some mitigation techniques and tips, which can help us in dealing with risks to some point;

**People switching jobs frequently:** This an external risk, which is not in control of an organization. People who provide estimates for the long term, if leave the job, can be reason for slipping on the estimates. So, it is better that estimation is done on small tasks or user stories, because if a person is leaving the company, it should not affect the others and future work on project.

**Unexpected/Unwanted events or holidays**: It is always better to provide the estimate in ranges because it gives you an idea how minimum and maximum time it can take to achieve a task or functionality. Keeping a space of buffer days in your plan can save you from slipping on schedule.

**Team members do not have clear idea about estimation techniques:** To have a clear idea and learning about estimation process and techniques, different workshops can be arranged in an organization give an idea about estimation techniques. This is would make a clear and combined understanding about estimation for everyone in the team and can help them in providing better estimates. Along with that, experiencing and learning from mistakes can always make them better estimators.

**Lack of guidance or missing role of software estimator**: if there would be a separate role of software estimator in an organization like Scrum master that would guarantee or produce better results for estimations. Like this, there would also be reinforcement of implementing estimation process under the organization.

**Extra demands from clients:** For extra work or change requests from clients, the commitments must be made separately; they should not be included in your existing or on-going plan. If there are extra demands from the client, communicate the existing plan to them and also include the risk, which would occur if they change the plan after handling change requests.

**Lack of user involvement:** User involvement is very crucial at any stage of project. Make sure that client is available at any point of confusion. Highlight the points in your plan where you think you need client's presence. If the client is not available to clear your points on required time, your timelines can get affected.

**Ambiguous requirements** It is not recommended to proceed on project with ambiguate requirements. Make sure you lock or finalize the requirements and everyone in the team is on same path in understanding of the requirements, only then your estimates can work.

**The less confidence on use of estimation process:** It should be made very clear to everyone in software development process that how software estimation plays an important part in project success factor. Once it is learned that, yes, software estimation is most important part of development life cycle, everyone would be part of it and make a positive change in the organizations or software houses.

**Lack of knowledge and use of estimation tool:** Use of estimation tool can help us refine the estimation process as most of our work is done with the help of the tool and the results are accurate too. The technical sessions or workshops can be arranged to give demo and knowledge of software estimation tools to the teams.

**Cultural Unfit:** While working with international teams Working with different cultural team can cause many problems or shifts. Such problem can also contribute to affect the estimate. For example, time zones are different and due to that the availability of the client or other team members can be affected, the communication gap can expand, also the holidays are different. So, we need to make sure and include risk mitigation plan accordingly while working with such teams.

**Availability of data:** Availability of industrial data would help us in accurate estimate and costing. Once we have data we can simulate and test our software. First you should encourage potential data providers to participate. Explain the amount of information they will bring to the project, and assure them that their information will be refined and used for the purposes in question. Also, if we have the access to the project historical data, it must also indicate to us that what kind of risks was introduced then, and what actions were taken to overcome or handle those risks. This can also reduce our work in identifying and mitigating the risk part in estimations.

To reduce risk, you need to develop a strategy to reduce your income. Meet with existing staff to look at the causes of profit (Poor working conditions, low wages, competitive labor market) and reduce those factors that are our cause before the start of the project. Once the work has started, money will be needed and strategies will be developed to ensure the continuity of the people as they move. You should also organize project teams so that the details of each development project are widely distributed. Standardization of documents is also positive step toward accurate costing. By assigning the backup staff member for every critical technologist. Estimation tools can be used to complete these initial reference and account frameworks for any uncertainties. After using initial estimation, managers can get a good and reliable idea about the size of their projects.

Whenever it comes to cost-effective software and estimation, common sense points to risk analysis. And yet most software project managers do it informally and automatically. For software projects the enemy is dangerous.

**VII. Results and Conclusion** In this paper, we analyzed that how important is to include the risks in your estimates while estimating the project plan, because risks are unavoidable part of the software development process. We mentioned that what are some risks that can occur in Pakistan Software Industry, and how they can affect the estimates.

| Software risks that can affect estimates | Risks that are universal | Risks in Pakistan Industry |
|---|---|---|
| People are switching jobs frequently | | ☑ |
| Unexpected/Unwanted events or holidays | | ☑ |
| Team members do not have clear idea about estimation techniques | | ☑ |
| Lack of guidance or missing role of software estimator | | ☑ |
| Extra demands from clients | ☑ | ☑ |
| Lack of user involvement | | ☑ |
| Ambiguous requirements | ☑ | ☑ |
| The less confidence on use of estimation process | | ☑ |
| Lack of knowledge and use of estimation tool | | ☑ |
| Cultural Difference | ☑ | ☑ |
| Availability of data | | ☑ |
| Unstable Environment | | ☑ |

Table 2 – The table concludes our idea of research, it shows the risks that are universal and the risks that are specific to Pakistan software industry.

However, there is no quick fix will make us better analysts and users of ratings. Effective evaluation occurs as a result of process and improvement information, education and training, good project management, use of appropriate tools and measurement techniques, adequate resources, and hard work. The process of integrating software risk into effort measurement activities begins with the anticipation and definition of software risk that may occur during software development, their testing, then the determination of the global project risk disclosure. This exposure to project risk is considered to be driving appropriate efforts as their impact on the development effort is proven using Pearson correlation testing.

# References

[1] B. Boehm, Software Engineering Economics. Englewood Cliffs, NJ: Prentice-Hall, 1981.

[2] K. Jantzen, K+K Jantzen Software Services GmbH,"Estimating the effects of project risks in software development projects",IWSM/MetriKon (2006)

[3] Pasha.org,"Annual Report of PASHA",2007-09.[Online].Available:https://www.pasha.org.pk/knowledge-center/industry-reports.[Accessed: 30-Jun-2020].

[4] June Verner, " What factors lead to software project failure ",(2008).

[5] J. J. G. Klein, "Software development risks to project efectiveness".

[6] The Standish Group Chaos Report, "https://www.standishgroup.com", 2015.

[7] Hooman Hoodat, "Classification and Analysis of Risks in Software Engineering", International Journal of Computer and Information Engineering, 2009.

[8] Softwaretestinghelp.com ,"Types Of Risks In Software Projects",2020. [Online].Available:https://www.softwaretestinghelp.com/types-of-risks-in-software-projects.[Accessed:20-Jan-2020].

[9] Tharwon Arnuphaptrairong ,"Top Ten Lists of Software Project Risks :Evidence from the Literature Survey",IMECS, (2014).

[10] Adanma Cecilia Eberendu,"Software Project Cost Estimation: Issues, Problems and Possible Solutions",International Journal of Engineering Science Invention,(2014).

[11] M. Uzzafer, " A contingency estimation model for software projects," *International Journal of Project Management,* 2013.

[12] F. W. R. L. S. L. Pfleeger, "Software Cost Estimation And Sizing Methods: Issues And Guidelines," 2005.

[13] R. S. Pressman, Softwaer Engineering – A Practicner's Approach,2010.

[14] O. T. a. O. J. Klakegg, "Challenges in Cost Estimation under Uncertainty—A Case Study of the Decommissioning of Barsebäck".

[15] J. G. Stellman, "Applied Software Project Management," 2005.

[16] A. &. G. J. Albrect, "Software function, source lines of code, and development effort prediction: A software science validation," *IEEE Transactions on Software Engineering,* (1983).

[17] D. Galorath, "Report on Failure of Software".

[18] B. W. Boehm, "Software risk management: principles and practices," *IEEE Software, 8, 32 41,*1991.

[19] J. R. a. K. Lyytinen, "Components of Software Development Risk:How to Address Them? A Project Manager Survey".

[20] Ali Javed,"Factors Affecting Software Cost Estimation in Developing Countries",I.J. Information Technology and Computer Science, 2013, 05